\documentclass[prl,aps,twocolumn,superscriptaddress]{revtex4-1}
\usepackage{bm}
\usepackage[colorlinks=true,linkcolor=blue,citecolor=blue]{hyperref}
\usepackage{times}
\usepackage{amsmath}
\usepackage{amssymb}
\usepackage{amsthm}
\usepackage{amsfonts}
\usepackage{enumerate}
\usepackage{latexsym}
\usepackage{ifpdf}
 \usepackage{psfrag}
\newcommand{\beq}{\begin{equation}}
\newcommand{\eeq}{\end{equation}}
\usepackage{graphicx}
\usepackage{makeidx}
\usepackage{color}

\usepackage{amsmath}
\usepackage{amsfonts}
\usepackage{amssymb}
\usepackage{graphicx}

\definecolor{amaranth}{rgb}{0.9, 0.17, 0.31}

\begin{document}

\title{Giant orbital polarization of Ni$^{2+}$ in square planar environment}

\author {Prithwijit Mandal}
\altaffiliation{Contributed equally}
\affiliation{Department of Physics, Indian Institute of Science, Bengaluru 560012, India}
\author {Ranjan Kumar Patel}
\altaffiliation{Contributed equally}
\affiliation{Department of Physics, Indian Institute of Science, Bengaluru 560012, India}
\author {Dibyata Rout}
\affiliation{Department of Physics, Indian Institute of Science Education and Research, Pune, Maharashtra 411008, India}
\author {Rajdeep Banerjee}
\affiliation{Solid State and Structural Chemistry Unit, Indian Institute of Science, Bengaluru 560012, India}
\author {Rabindranath Bag}
\affiliation{Department of Physics, Indian Institute of Science Education and Research, Pune, Maharashtra 411008, India}
\author {Koushik Karmakar}
\affiliation{Department of Physics, Indian Institute of Science Education and Research, Pune, Maharashtra 411008, India}
\author {Awadhesh Narayan}
\affiliation{Solid State and Structural Chemistry Unit, Indian Institute of Science, Bengaluru 560012, India}
\author {John W. Freeland}
\affiliation{Advanced Photon Source, Argonne National Laboratory, Argonne, Illinois 60439, USA}
\author {Surjeet Singh}
\affiliation{Department of Physics, Indian Institute of Science Education and Research, Pune, Maharashtra 411008, India}
\author {Srimanta Middey}
\email{Corresponding author. Email: smiddey@iisc.ac.in }
\affiliation{Department of Physics, Indian Institute of Science, Bengaluru 560012, India}

\begin{abstract}
{Understanding the electronic behavior of Ni$^{2+}$ in a square planar environment of oxygen is the key to unravel the origin of the recently discovered superconductivity in the hole doped nickelate Nd$_{0.8}$Sr$_{0.2}$NiO$_2$. To identify the major similarities/dissimilarities between nickelate and cuprate superconductivity, the study of the electronic structure of Ni$^{2+}$ and Cu$^{2+}$ in an identical square planar environment is essential. In order to address these questions,  we investigate the electronic structure of Sr$_2$CuO$_3$  and Ni doped Sr$_2$CuO$_3$ single crystals containing (Cu/Ni)O$_4$ square planar units. Our polarization dependent X-ray absorption spectroscopy experiments for Ni in Sr$_2$Cu$_{0.9}$Ni$_{0.1}$O$_3$ have revealed very large orbital polarization, which is a characteristic feature of high $T_c$ cuprate.  This arises due to  the low spin $S$=0 configuration with two holes in Ni 3$d_{x^2-y^2}$ orbitals -  in contrast to the expected high spin  $S$=1 state from Hund's first rule. The presence of such $S$=0 Ni$^{2+}$ in hole doped nickelate would be analogous to the Zhang Rice singlet.  However, the Mott Hubbard insulating nature of the NiO$_4$ unit would point towards a different  electronic phase space of  nickelates, compared to high $T_c$ cuprates.}
\end{abstract}
 
\maketitle

{\bf \small \color{amaranth}INTRODUCTION} 

 Ever since the discovery,  understanding  the  mechanism  of high temperature superconductivity in cuprates remains an open  problem in condensed matter physics~\cite{Bednorz:1986p189}. Over the years, the importance of several  factors, such as the antiferromagnetic, insulating phase of the parent compound, large orbital polarization, strong hybridization between Cu $d$ and O $p$ states,   and Zhang Rice  (ZR) physics have been recognized  to play a role in the hole doped superconducting phase ~\cite{Lee:2006p17,Fradkin:2015p457,Keimer:2015p179}. After the prediction of achieving cuprate like Fermi surface in a three-dimensional compound LaNiO$_3$ through orbital engineering~\cite{Anisimov:1999p7901,Chaloupka:2008p016404,Hansmann:2009p016401}, various types of artificial structures of $RE$NiO$_3$ family have been extensively investigated~\cite{Middey:2016p305,Freeland:2011p57004,Benckiser:2011p189,Wu:2013p125124}. However, the observed orbital polarization in an octahedral crystal field  is  low (maximum 25\%) and superconductivity remains elusive there, prompting to search for other Ni based compounds with lower dimensional structure~\cite{Poltavets:2010p206403,Cheng:2012p236403,Zhang:2017p864}.  
 
\begin{figure}
	\centering{
		\includegraphics[scale=0.35]{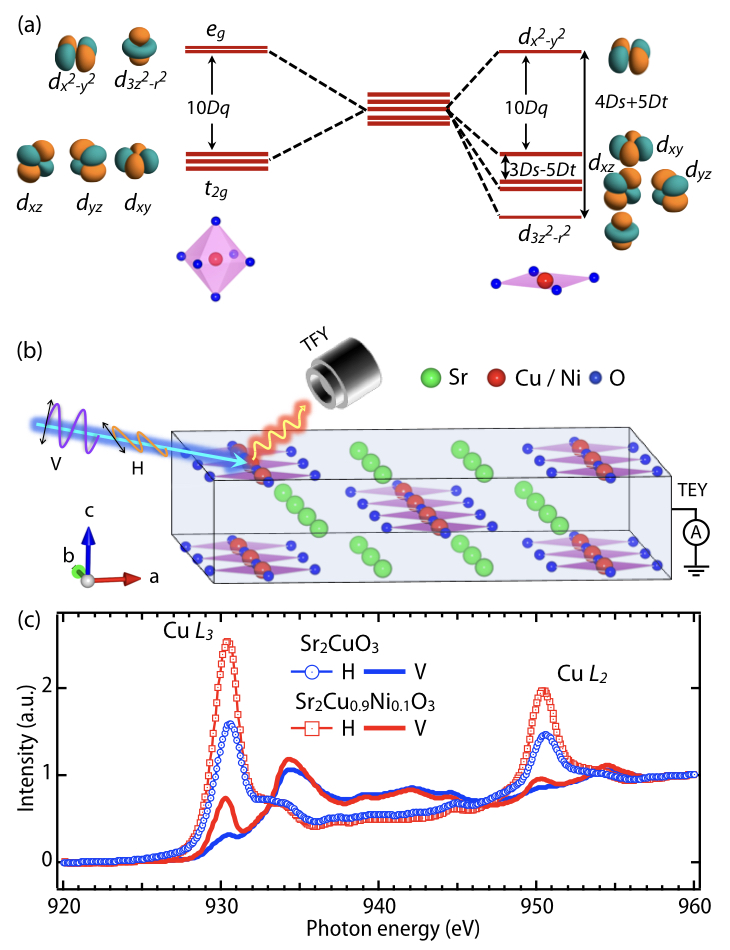}
		\caption{ (a) Energy level splitting of transition metal $d$ orbitals  due to point-charge (Coulomb) contributions under octahedral ($O_h$) and square planar ($D_{4h}$) environment~\cite{Moretti_Sala_2011}.  The ordering of these levels can be modified in solids due to the hybridization between transition metal $d$ and oxygen $p$ states~\cite{Ushakov:2011p445601}. (b) Crystal structure of Sr$_2$(Cu,Ni)O$_3$ and experimental arrangement for  XAS experiment with vertically (V) and horizontally (H) polarized X-ray.  TFY and TEY denote total fluorescence yield and total electron yield, respectively. (c) Cu $L_3$-edge XAS recorded in bulk sensitive TFY mode with in-plane and out-of-plane polarized light for Sr$_2$CuO$_3$ and Sr$_2$Cu$_{0.9}$Ni$_{0.1}$O$_3$. }
		\label{Fig1}}
\end{figure}

 The recent finding of superconductivity in Sr doped NdNiO$_2$ having Ni in a square planar oxygen environment  is a monumental development~\cite{Li:2019p624}, and  strongly implies that the local structure ($O_h$ vs. $D_{4h}$ in Fig.~\ref{Fig1}(a)) plays a crucial role.  This discovery has also led to a large number of works within a short span of time to understand the origin and nature of the superconducting phase~\cite{Hepting:2020p381,li2020superconducting,zeng:2020phase,Botana:2020p011024,Lechermann:2020p081110,Jiang:2019p201106,karp:2020,Hu:2019p032046,adhikary2020orbital,Mi:2020p207004,Hu:2019p032046,Zhang2020p023112,Zhang:2020p013214,Lang:2020doped,karp:2020many,Krishna:2020arxiv}.  While the parent members of the superconducting nickelate and cuprate family are isoelectronic ($d^9$),  several key differences have been  noted: unlike  the antiferromagnetic insulating phase of the cuprate,  NdNiO$_2$ is weakly conducting and nonmagnetic~\cite{Li:2019p624}.  Ni $d_{x^2-y^2}$ states are  found to be strongly hybridized with Nd 5$d$ states~\cite{Hepting:2020p381} and the superconducting $T_c$ of infinite layer nickelates is much smaller compared to the analogous hole doped CaCuO$_2$~\cite{li2020superconducting,zeng:2020phase,Azuma:1992p775}.  There is also an ongoing strong  debate about the location  of doped holes  (O $p$ vs. Ni $d$) and  the nature of Ni spins states (singlet vs. triplet), which would be intimately connected with the orbital configuration and Hund's coupling~\cite{Lang:2020doped,karp:2020many,Mi:2020p207004,Hu:2019p032046,Zhang2020p023112,Zhang:2020p013214,goodge2020doping}.    However, the actual scenario is unknown due to  the very limited amount of experimental literature available.  This prompts us to probe other compounds containing both  Cu and Ni in four fold coordinated environment of oxygen.

 Sr$_2$CuO$_3$ (SCO) contains square planar CuO$_4$ polyhedral units, which form  linear chains along the $b$ axis, as shown in Fig.~\ref{Fig1}(b)~\cite{Motoyama:1996p3212}. This quasi one dimensional system undergoes antiferromagnetic ordering around 5 K~\cite{Kojima:1997p1787}, and  exhibits several intriguing phenomena, including  spin-charge separation~\cite{Neudert:1998p657}, spin-orbital separation~\cite{Schlappa:2012p82}, and spin Seebeck effect~\cite{Hirobe:2017p30}.  Hole doping in SCO by incorporating excess oxygen further results in superconductivity~\cite{Hiroi:1993p315}.   In addition, one can replace a small fraction of Cu of SCO with Ni without modifying structure~\cite{Karmakar:2015p224401}, providing us a unique platform to compare the  electronic structure of Cu and Ni with a square planar environment.  

 The electronic properties of transition metal (TM) oxides are  described  very often within the scheme of well-known Zaanen-Sawatzky-Allen (ZSA) phase diagram~\cite{Zaanen:1985p418,Nimkar:1993p7355}, which considers on-site Coulomb repulsion at TM site ($U$), TM-to-O charge transfer energy ($\Delta$) and the electronic hopping interaction strength ($t$). Cu in SCO has an extremely small value of $\Delta$~\cite{Maiti:1998p1572}, implying that the ground state of Cu$^{2+}$ in SCO consists of a linear combination of $d^9$ and dominant $d^{10}\underline{L}$ configurations ($\underline{L}$ denotes a hole in O 2$p$ state) -  contrary to the pure  $d^9$ state expected from an  ionic picture~\cite{walters:2009p867}.

 In this work, we have investigated the electronic and magnetic structure of Sr$_2$CuO$_3$ (SCO) and Sr$_2$Cu$_{0.9}$Ni$_{0.1}$O$_3$ (SCNO) single crystals by magnetic measurement, synchrotron based X-ray linear dichroism (XLD) measurements, density functional theory (DFT) and cluster calculations.  We observed extremely large orbital polarization of Ni in   Sr$_2$Cu$_{0.9}$Ni$_{0.1}$O$_3$ single crystal, which arises due to  the low spin $S$=0 configuration  with completely unoccupied 3$d_{x^2-y^2}$ orbitals of Ni$^{2+}$. Ni doping of SCO also enhances the contribution of the $d^9$ configuration in the ground state of Cu$^{2+}$.   Our comprehensive study has also affirmed Mott Hubbard insulating nature of the NiO$_4$  unit and larger charge transfer energy compared to high $T_c$ cuprate.

{\bf \small \color{amaranth}RESULTS}

Single crystals of SCO and SCNO, grown by  the traveling solvent floating zone technique~\cite{karmakar:2015p4843} have been investigated in this study.  We first discuss the results of polarization dependent XAS (X-ray absorption spectroscopy) experiment on Cu $L_{3,2}$ edges, which is known to be a very powerful technique  to probe orbital symmetry of cuprates~\cite{Nucker:1995p8529,Knupfer:1997pR7291,Neudert:2000p10752,Chakhalian:2007p1114}.   In Fig.~\ref{Fig1}(c), we compare linear polarization dependent XAS of Cu $L_3$ and $L_2$ edges for  SCO and SCNO.  XAS  (I$_\mathrm{ab}$)   of SCO recorded with in-plane (H) polarized light consists of sharp peaks around 930.5 eV and 950.5 eV, which are associated with the transition from the spin-orbit split Cu $2p_{3/2}$ and  Cu $2p_{1/2}$ core levels, respectively, and the final state configuration is 2$\underline{p}$3$d^{10}$ (2$\underline{p}$ denotes a core hole in Cu 2$p$ states). The much lower intensity of these peaks in XAS (I$_\mathrm{c}$) with the out-of-plane (V) polarization signifies that the upper Hubbard band (UHB) is predominately contributed by Cu $d_{x^2-y^2}$ orbitals, which is expected for a $d^9$  electronic configuration in $D_{4h}$ symmetry.  The features around 934 eV and 954.5 eV, which are much stronger for the out-of-plane polarization, are related to the transitions into Cu 3$d_{3z^2-r^2}$ derived states, which become partially unoccupied due to the hybridization with Cu 4$s$ states~\cite{Neudert:2000p10752}.  Upon Ni doping, the intensity of the white line increases significantly for both polarizations  though the line shape remains unchanged (Fig.~\ref{Fig1}(c)).  The  origin of the increase of the $d^9$ configuration upon Ni doping will be discussed later in this paper.

\begin{figure}
	\centering{
		\includegraphics[scale=0.48]{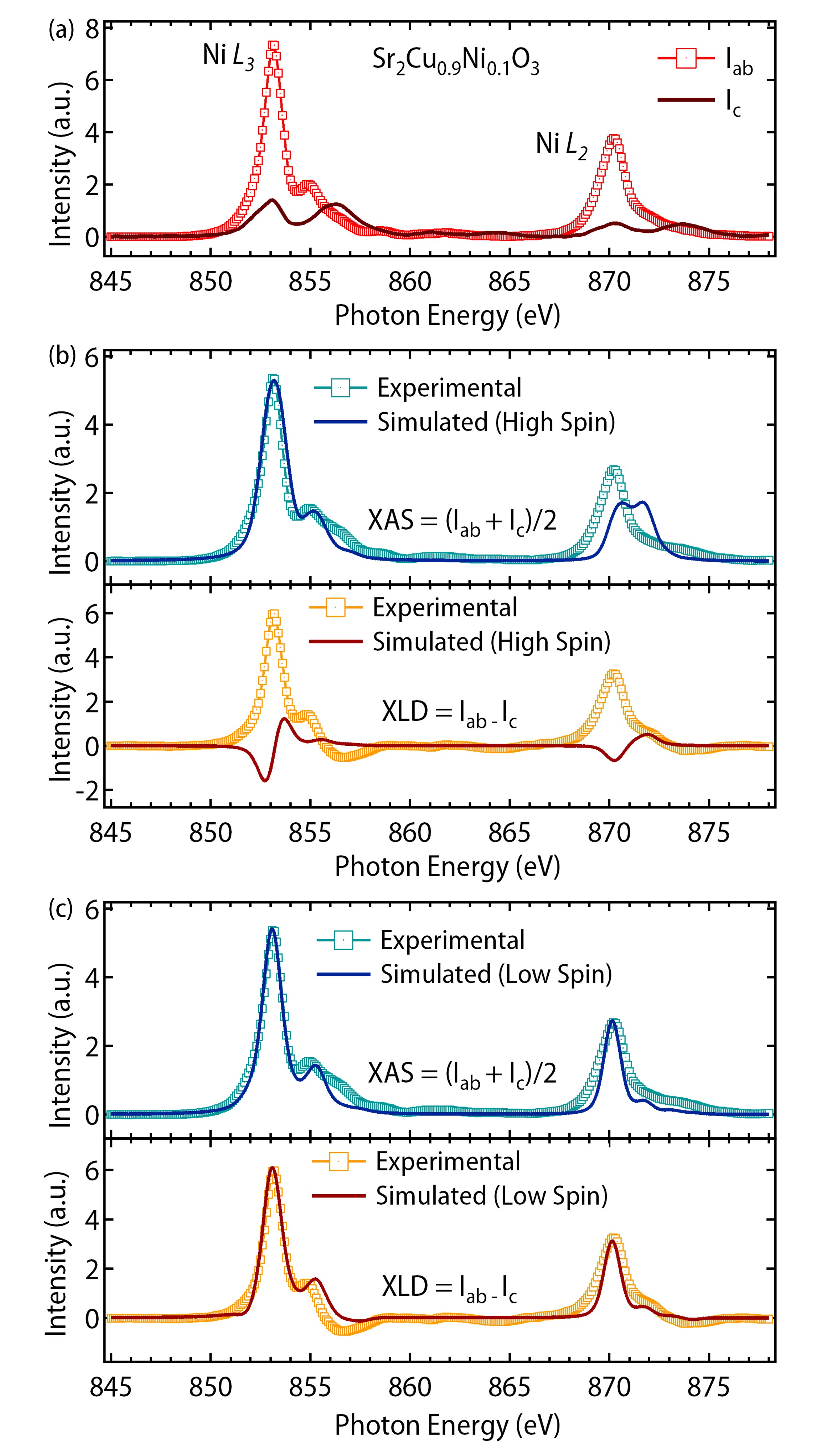}
		\caption{ (a) Ni $L_3$ and $L_2$ XAS (TFY mode) for Sr$_2$Cu$_{0.9}$Ni$_{0.1}$O$_3$, recorded with V and H polarization. These spectra have been plotted after subtracting the background using a step function (see Supplementary Materials) (b) Top panel: Experimentally observed isotropic XAS and the simulated XAS spectra. Bottom panel: Experimentally observed and simulated XLD. The simulation is done for Ni$^{2+}$ in high spin (S = 1) configuration. (c) shows the same for Ni$^{2+}$ in low spin ($S$ = 0) configuration.  The small feature around 856 eV is not captured by the simulation since the 4$s$ orbitals were not included in the basis set of cluster calculations. } \label{Fig_2}}
\end{figure}

 Next, we have performed similar polarization dependent XAS experiments at Ni \textit{L$_{3,2}$} edges  of SCNO at 300 K. As shown in Fig. \ref{Fig_2}(a), the two intense peaks around 853 eV \& 870 eV in the spectra  (I$_\mathrm{ab}$)  recorded with the in-plane polarization, correspond to the transitions from Ni $2p_{3/2}$ and Ni $2p_{1/2}$ core levels respectively. The spectrum   (I$_\mathrm{c}$)  recorded with the out-plane polarization shows significantly low intensity indicating that the holes predominantly occupy the $d_{x^2 - y^2}$ bands. The peaks around 856 eV and 874 eV, show higher intensity in   I$_\mathrm{c}$   than   I$_\mathrm{ab}$ . Similar to the case of Cu, these features are related to the hybridization of Ni 3$d_{3z^2 - r^2}$ and Ni 4$s$ states. Since the experimental realization of  large orbital polarization (OP) in Ni based oxides,  with important implications for superconductivity, has been a topic of paramount interest~\cite{Disa:2015p026801,Wu:2013p125124,Zhang:2017p864,Liao_2019}, we have further evaluated OP of Ni in the present case. According to the sum rule of linear polarization, the ratio of holes in $d_{3z^2 - r^2}$ and $d_{x^2 - y^2}$~\cite{Wu:2013p125124} is defined as
\begin{equation}
X =\frac{ h_{3z^2 - r^2}} {h_{x^2 - y^2}}=\frac{3\mathrm{I'_c}}{4\mathrm{I'_{ab}-I_{c}'}} 	
\end{equation}
with $ \mathrm{I'_{ab,c}}$=$\int \mathrm{I_{ab,c}}(E)dE$.  	By this definition, $X$=1 corresponds to equal hole population in  $d_{3z^2 - r^2}$ and $d_{x^2 - y^2}$ orbitals and $X$= 0 signifies 100\% $d_{x^2-y^2}$ character of holes.  We have obtained $X\approx$  0.23 (also see Supplementary Materials) in the present case. This implies 81\% of the holes occupy Ni 3$d_{x^2 - y^2}$ orbital, which is the highest among the existing literature of Ni based complex oxides (see Table ~\ref{T1}).

 	  \begin{table}[h]
	 \caption{\label{T1} Comparison of hole ratio  for  Ni 3$d_{x^2-y^2}$ and Ni   3$d_{3z^2-r^2}$ orbitals and \% of holes in Ni 3$d_{x^2-y^2}$ for Sr$_2$Cu$_{0.9}$Ni$_{0.1}$O$_3$ with  literature.}
		\centering
\begin{tabular}{lcccc}
		\hline
		\hline
		System&  & $X$   & \% of holes in Ni $d_{x^2 - y^2}$ & Reference\\
	\hline
	SrCuO$_2$/LaNiO$_3$  & & 0.7 & 59\% & \onlinecite{Liao_2019}\\
	LaTiO$_3$/LaNiO$_3$/LaAlO$_3$  & & 0.55 & 65\% & \onlinecite{Disa:2015p026801} \\
	La$_4$Ni$_3$O$_8$ & & 0.4-0.5& 67-71\%&\onlinecite{Zhang:2017p864} \\
	Sr$_2$Cu$_{0.9}$Ni$_{0.1}$O$_3$ & & 0.23 & 81\%& present work\\
	 
	 	\hline
	\hline
		\end{tabular}
		\end{table}

  The orbital polarization ($P$) is defined in the literature in terms of the electronic occupation as ~\cite{Wu:2013p125124}
\begin{equation}
P=\left|\frac{n_{x^2 - y^2}-n_{3z^2 - r^2}}{n_{x^2 - y^2}+n_{3z^2 - r^2}}\right|=\left|\left( \frac{4}{n'}-1\right)\frac{(X-1)}{(X+1)}\right|
\end{equation}
where $n_{x^2 - y^2}$= 2 - $h_{x^2 - y^2}$ and $n_{3z^2 - r^2}$=2 - $h_{3z^2 - r^2}$ are the number of electrons in $d_{x^2 - y^2}$ and $d_{3z^2 - r^2}$ orbitals, respectively.  It is important to note that the value of $n'$ (=$n_{x^2 - y^2}$ +  $n_{3z^2 - r^2}$) can not be determined unambiguously  for many compounds due to the presence of charge transfer from O 2$p$ to Ni 3$d$ orbitals~\cite{Wu:2013p125124,Peil:2014p045128} . In the present case,  $n'$ =2 as the ground state wavefunction is almost entirely contributed by the $d^8$ configuration (discussed in the later part of the paper). We have obtained $P\approx$ 63\% for the SCNO, which is also larger compared to all existing literatures on Ni based complex oxides~\cite{Middey:2016p305,Disa:2015p026801,Wu:2013p125124,Middey:2016p056801,Zhang:2017p864,Liao_2019}.

In order to understand the origin of this unprecedentedly large orbital polarization, we have calculated  XAS [(I$_\mathrm{ab}$ + I$_\mathrm{c}$)/2] and XLD [I$_\mathrm{ab}$-I$_\mathrm{c}$] of Ni $L_{3,2}$ edges considering a NiO$_4$ cluster with $D_{4h}$ symmetry and Ni$^{2+}$ ionic configuration  using CTM4XAS~\cite{Stavitski:2010p687} and Quanty ~\cite{retegan_crispy} programs that incorporate   ligand field multiplet theory.  First, we have simulated the XAS  and XLD  spectra  (Fig. \ref{Fig_2}(b)) for the high spin $S$ = 1  configuration ~\cite{Utz_2017,van_der_Laan_1988} with $U$= 7 eV and $\Delta$ = 9 eV (all other parameters, used for the simulations have been listed in Ref.~\onlinecite{calc}).  Clearly, neither the simulated XAS nor the XLD matches with the corresponding experimental spectra.  In particular, the experimentally observed $L_2$-edge XAS shows only one peak around 870 eV whereas the simulated spectra shows two peaks.  The shape of the simulated XLD is completely different and the amount of anisotropy is significantly lower compared to the experimental data.  We have also simulated XAS and XLD with the  same values of $U$ and $\Delta$ but with low spin configuration  ($S$ = 0)  by adjusting crystal field parameters~\cite{calc}, which ensures the following  expected crystal field splitting $d_{3z^2 - r^2}<d_{xz/yz}<d_{xy}<d_{x^2 - y^2}$ for a square planar environment~\cite{Moretti_Sala_2011}.  The excellent matching (Fig. ~\ref{Fig_2}(c)) between the experimentally observed and simulated spectra with  low spin configuration  establishes  Ni$^{2+}$ is  in  $S$=0 state in the square planar environment.  The ground state  of Ni$^{2+}$ is  strongly dominated by the $| d^8\rangle$ configuration  ($\approx$  0.97 $| d^8\rangle$ + 0.03 $| d^9\underline{\textit{L}}\rangle$). Simulated spectra with $U > \Delta$ also do not match with experimental observation (see Supplementary Materials), implying electronically NiO$_4$ belongs to the Mott Hubbard insulating region  of ZSA phase diagram~\cite{Zaanen:1985p418} - strikingly similar to the recent experimental reports of the hole doped NdNiO$_2$~\cite{Hepting:2020p381,goodge2020doping}.  Due to the large $\Delta$, Ni $d$ is less hybridized with O $p$, compared to the Cu $d$-O $p$ hybridization. Thus, the replacement of a fraction of Cu by Ni pushes holes from O $p$ towards Cu $d$ around the dopant Ni. This enhances the relative contribution of $d^9$ configuration (also confirmed by magnetic measurement, discussed later in the paper), resulting in the  observed strong increase in  the white line intensity in Cu XAS upon Ni doping (Fig.~\ref{Fig1}(c)).

\begin{figure}
	\centering{
		\includegraphics[scale=0.54]{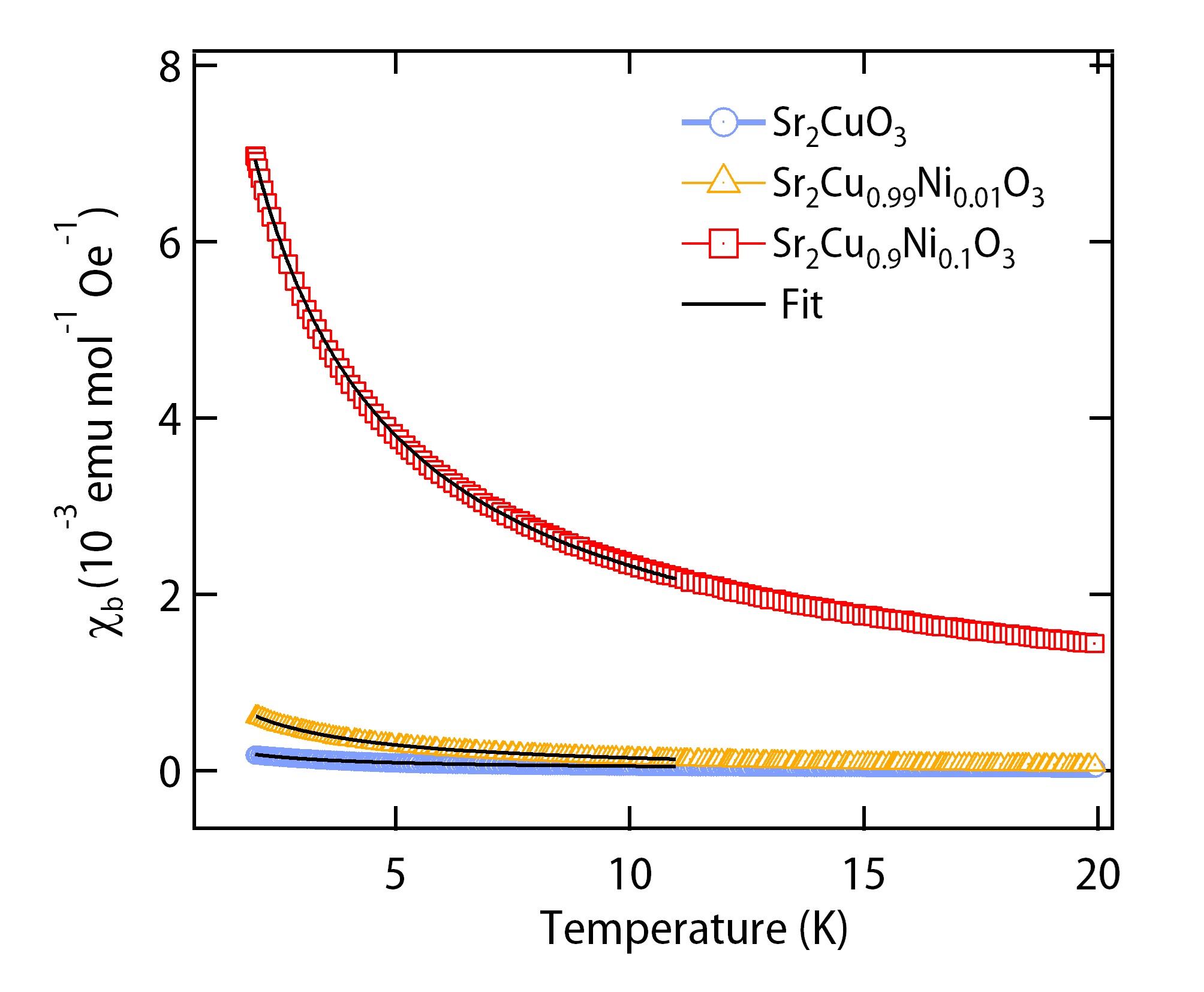}
		\caption{  Temperature dependence of magnetic susceptibility ($\chi$) and fitting by Curie-Weiss  (CW) law.   In principle, depending on the temperature range, one should also include not only the chain susceptibility, i.e., the susceptibility of an endless antiferromagnetic chain~\cite{Johnston:2000p9558}, but also the boundary contribution arising from the staggered moment at the free ends of a chain segment~\cite{Fujimoto:2004p037206,Karmakar:2015p224401}. However as shown by Sirker et al., the generalized expression of $\chi$, including all the terms mentioned above, reduces to the CW equation in the low-temperature limit~\cite{Sirker:2007p137205}. Hence, we used the data below 10 K and fitted that to CW equation for obtaining the effective moment from which the spin state of Ni can be inferred. The data for 1\% Ni doped SCO sample has been adapted from Ref.~\onlinecite{Karmakar:2015p224401}.} \label{Figmag}}
\end{figure}

The increase of the $d^9$ component of Cu due to the presence of nonmagnetic Ni$^{2+}$ is also evident in bulk magnetic measurements. The low-temperature magnetic susceptibility ($\chi$) plots of pristine, 1\%, and 10\% Ni doped SCO samples are shown in Fig. ~\ref{Figmag}. As expected,  $\chi$ for the pristine sample is very small due to the antiferromagnetic coupling between the neighboring Cu spins. Upon doping with Ni at the Cu-site, the -O-Cu-O-Cu-O- chain breaks into finite length segments due to an effective spin 0 on the Ni ion. The number of segments thus produced being $p$ + 1 $\sim  p$ for large $p$, where $p$ is the concentration of Ni in the sample. Assuming a statistically random distribution of Ni ions over the chain length, it is reasonable to consider that one-half of the chain segments will have an odd number of Cu spins, and even for the remaining. Since Cu-spins are antiferromagnetically paired,  the net moment on the even-length segments   tends to 0 at sufficiently low-temperatures~\cite{Ami:1995p5994}. In contrast, the odd-length segments   remain with an uncompensated Cu spin, which   results in an enhancement in the  magnetic response of Ni doped samples,  as seen in our measurement (Fig. ~\ref{Figmag}). A very weak Curie-tail in the pristine sample can be attributed to the presence of intrinsic defects.  We fitted the low-temperature susceptibility of our samples using the Curie-Weiss (CW) law: $\chi$=$\chi_0$+  $C/(T- \theta_{CW} )$ where $\chi_0$ is the temperature independent contribution arising due to Van Vleck paramagnetism and core diamagnetism, $C$ is the Curie-constant, and $\theta_{CW}$ is the Weiss temperature. The best fit parameters for the three samples are tabulated in Table~\ref{T2}.

 	  \begin{table}
	 \caption{\label{T2}The best-fit values of the fitting parameters using the CW equation.}
		\centering
\begin{tabular}{lccccc}
		\hline
		\hline
		Doping & $\chi_0$ (emu &  $C$  (emu & $\theta_{CW}$ & $\mu_\mathrm{eff}$ & $\mu_\mathrm{eff}^{odd}$\\
	($p$) &  mol$^{-1}$Oe$^{-1}$) &  mol$^{-1}$Oe$^{-1}$ K) & (K) &  ($\mu_B$/f.u.) & ($\mu_B$/1/2 Ni)\\
	\hline
	0 & $\sim$ -5 $\times$ 10$^{-5}$ &  0.0005(4) & -0.6 & $\sim$ 0.06 & - \\
	0.01 & $\sim$ -3 $\times$ 10$^{-5}$ & 0.0018(2) & -1(1)  &$\sim$ 0.12  & $\sim$ 1.74  \\
	0.1 & $\sim$ 5 $\times$ 10$^{-4}$ & 0.018(2) & -2(1) & $\sim$ 0.38  & $\sim$ 1.70 \\
	\hline
	\hline
		\end{tabular}
		\end{table}

\begin{figure*}
	\centering{
		\includegraphics[scale=0.38]{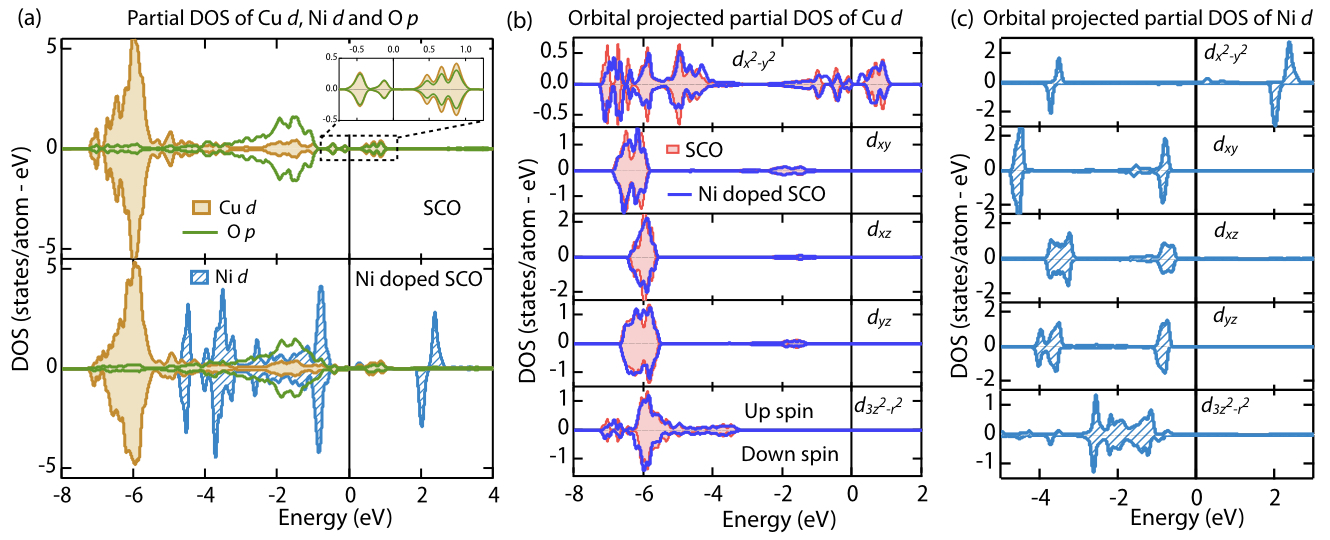}
		\caption{ Orbital projected partial density of states: (a) comparison of DOS of Cu-d, O-p states of undoped SCO (upper panel) with DOS of Cu-d, O-p and Ni-d states of Ni doped SCO (lower panel), (b) comparison of different Cu d-states of doped (line) and undoped (line + shaded) SCO, (c) different d- orbital projected DOS of Ni in doped SCO. Note that the vertical line at zero energy indicates the Fermi level.} \label{Fig4}}
\end{figure*}

It is reassuring to note that the value of   $C$ increases by a factor of 10 on increasing the Ni concentration from $p$ = 0.01 to 0.1. $\chi_0$  for SCNOis positive and about an order of magnitude higher than for SCO. Such an enhancement is likely due   to the Van Vleck paramagnetism associated with   unoccupied $d_{x^2-y^2}$ orbital of Ni$^{2+}$ ion in the low-spin state~\cite{van1932book}. From $C$, the $\mu_\mathrm{eff}$ (=$\sqrt{8C}$) for the two Ni doped samples ($p$ = 0.01 and 0.1) turned out to be $\sim$ 0.12 $\mu_B$ and $\sim$ 0.38 $\mu_B$ per formula unit respectively.  The average effective moment on the odd-length segments (=$\sqrt{8C/(0.5p)}$) turns out to be 1.74  $\mu_B$ and 1.70 $\mu_B$, respectively for the two samples. These values are close to an effective moment of $\sim$  1.73 $\mu_B$ expected for a $S$ = 1/2 ($d^9$) system, confirming our conclusion of Cu XAS measurement.

 To obtain a microscopic understanding of our experimental observations, we have carried out first-principles density functional theory (DFT) calculations for Sr$_2$CuO$_3$ and Sr$_2$Cu$_{0.875}$Ni$_{0.125}$O$_3$ [see Methods for details of the calculations]. The calculated orbital projected density of states (DOS) are shown in Fig.~\ref{Fig4}.  In contrast with  the transition metal oxides with positive $\Delta$,  the bonding states of SCO are primarily contributed by  Cu $d$ orbitals due to negative $\Delta$~\cite{Ushakov:2011p445601}.  The Cu $d_{xy}$, $d_{xz}$, $d_{yz}$ and $d_{3z^2-r^2}$ bonding states are spread around 6 eV below the Fermi level (Fig.~\ref{Fig4}(b)).  The valence band and conduction band edge of undoped SCO comes predominantly from O $p$ states (inset of upper panel of Fig.~\ref{Fig4}(a)),  as expected for the covalent insulating nature of SCO~\cite{Maiti:1998p1572}.   The $d_{x^2-y^2}$ antibonding states are partially unoccupied and show a pronounced contribution to the upper Hubbard band, in line with our XAS results. We find that the magnetic moment on Cu is close to 0.19 $\mu_B$.  The slight overestimation of this moment, compared to the experimentally observed moment  ($\sim$ 0.06$ \mu_B$)~\cite{Kojima:1997p1787} is likely to be related to the presence of the quantum fluctuation effects in SCO, which are not accounted for in a mean field approach like DFT..  The much smaller moment, compared to the expected moment of 1.73$\mu_B$ for a  pure $d^9$ ($S$=1/2) configuration,  again signifies that the ground state of Cu$^{2+}$ in SCO is predominantly $d^{10}\underline{L}$ configuration with small admixture with $d^9$. In Ni doped SCO, there is a small increase in Cu magnetic moments, which range from 0.19 to 0.25 $\mu_B$ and a net moment of $\sim$ 0.13 $\mu_B$ per formulae unit. This also supports our observation about the increase of  white line intensity of Cu XAS (Fig.~\ref{Fig1}(c)) and magnetic measurement i.e. the enhancement of $d^9$ configuration of Cu upon Ni doping. For Ni doped SCO, the contribution arising from different Ni $d$ orbitals is shown in Fig. ~\ref{Fig4}(c) (also see lower panel of Fig. ~\ref{Fig4}(a)).  We find that the antibonding states of Ni $d_{x^2-y^2}$ orbital are completely unoccupied, while all   other Ni 3$d$ states are completely  filled. This low spin $S$=0 configuration of Ni with holes in the   $d_{x^2-y^2}$ orbital is in excellent agreement with our experimental observations.

 {\bf \small \color{amaranth}DISCUSSION} 

To summarize, our element sensitive X-ray absorption spectroscopy experiments with linear polarized light on Sr$_2$Cu$_{0.9}$Ni$_{0.1}$O$_3$  have found that 81\% holes of Ni$^{2+}$ sites occupy the $d_{x^2-y^2}$ orbital, which is the highest among all Ni containing oxides reported so far. Ni doping in Sr$_2$CuO$_3$  also results in hole transfer from O $p$ to Cu $d$ orbitals. Cluster calculations  have not only confirmed the Mott Hubbard insulating nature of NiO$_4$  units but also  revealed that the observed giant orbital polarization arises due to low spin ($S$=0)  configuration of Ni$^{2+}$. This nonmagnetic state of Ni$^{2+}$ has been further corroborated by $ab$ $initio$ theory.  

Finally, we also note that our investigated system has several parallels with the recently discovered superconducting nickelates.  Our system comprises of NiO$_4$ units forming one dimensional chains, while the superconducting nickelates have a two dimensional NiO$_2$ plane. It is expected that the increase in the dimensionality of the system would further enhance $\Delta$~\cite{Maiti:1999p12457}, and therefore our conclusions about the Mott-Hubbard nature and the low spin configuration of Ni$^{2+}$ should also hold for nickelate superconductors. The larger charge transfer energy might be responsible for the observed lower $T_c$ in hole doped nickelates~\cite{Zhang:2020p013214,Lechermann:2020p081110,Jiang:2019p201106}.

{\bf \small \color{amaranth}MATERIALS AND METHODS} 

Single crystals of SCO and SCNO were grown under O$_2$ atmosphere using the traveling solvent floating zone technique as reported elsewhere~\cite{karmakar:2015p4843}.  All the experiments reported here are carried out on as grown crystals.  The magnetic susceptibility  was measured using a Physical Property Measurement System, Quantum Design, USA. X-ray absorption spectra (XAS) with vertically (V) and horizontally (H) polarized X-ray were recorded at room temperature in 4-ID-C beamline of Advanced Photon Source, Argonne National Laboratory, USA.

All theoretical calculations have been performed within the framework of density functional theory as implemented in the Quantum Espresso package~\cite{Giannozzi:2009p395502}. The wave functions are expanded in a plane wave basis set with kinetic and charge density cut offs set to 45 Ry and 450 Ry respectively. The exchange-correlation interactions are taken into account by the Perdew-Burke-Ernzeroff form of generalized gradient approximation~\cite{Perdew:1996p3865}. The ion-electron interactions are described by ultrasoft pseudopotentials~\cite{Vanderbilt:1990p7892}. The on-site Coulomb interaction is modeled using LDA+$U$ approach in the linear-response method of Cococcioni and de Gironcoli~\cite{Cococcioni:2005p3}, with $U$ = 5 eV and 9 eV on Ni and Cu, respectively.


 

\begin{thebibliography}{10}

\bibitem{Bednorz:1986p189}
J.~G. Bednorz, K.~A. Muller, Possible high {$T_c$} superconductivity in the
  {Ba-La-Cu-O} system.
\newblock {\it Zeitschrift fur Physik B Condensed Matter\/} {\bf 64}, 189--193
  (1986).

\bibitem{Lee:2006p17}
P.~A. Lee, N.~Nagaosa, X.-G. Wen, Doping a mott insulator: Physics of
  high-temperature superconductivity.
\newblock {\it Rev. Mod. Phys.\/} {\bf 78}, 17--85 (2006).

\bibitem{Fradkin:2015p457}
E.~Fradkin, S.~A. Kivelson, J.~M. Tranquada, Colloquium: Theory of intertwined
  orders in high temperature superconductors.
\newblock {\it Rev. Mod. Phys.\/} {\bf 87}, 457--482 (2015).

\bibitem{Keimer:2015p179}
B.~Keimer, S.~A. Kivelson, M.~R. Norman, S.~Uchida, J.~Zaanen, From quantum
  matter to high-temperature superconductivity in copper oxides.
\newblock {\it Nature\/} {\bf 518}, 179--186 (2015).

\bibitem{Anisimov:1999p7901}
V.~I. Anisimov, D.~Bukhvalov, T.~M. Rice, Electronic structure of possible
  nickelate analogs to the cuprates.
\newblock {\it Phys. Rev. B\/} {\bf 59}, 7901--7906 (1999).

\bibitem{Chaloupka:2008p016404}
J.~c.~v. Chaloupka, G.~Khaliullin, Orbital order and possible superconductivity
  in {L}a{N}i{O}$_{3}$/{L}a{MO}$_{3}$ superlattices.
\newblock {\it Phys. Rev. Lett.\/} {\bf 100}, 016404 (2008).

\bibitem{Hansmann:2009p016401}
P.~Hansmann, X.~Yang, A.~Toschi, G.~Khaliullin, O.~K. Andersen, K.~Held,
  Turning a nickelate fermi surface into a cupratelike one through
  heterostructuring.
\newblock {\it Phys. Rev. Lett.\/} {\bf 103}, 016401 (2009).

\bibitem{Middey:2016p305}
S.~Middey, J.~Chakhalian, P.~Mahadevan, J.~W. Freeland, A.~J. Millis, D.~D.
  Sarma, Physics of ultrathin films and heterostructures of rare-earth
  nickelates.
\newblock {\it Annual Review of Materials Research\/} {\bf 46}, 305--334
  (2016).

\bibitem{Freeland:2011p57004}
J.~W. Freeland, J.~Liu, M.~Kareev, B.~Gray, J.~W. Kim, P.~J. Ryan,
  R.~Pentcheva, J.~Chakhalian, Orbital control in strained ultra-thin
  {L}a{N}i{O}$_{3}$/{L}a{A}l{O}$_{3}$ superlattices.
\newblock {\it Europhysics Letters\/} {\bf 96}, 57004 (2011).

\bibitem{Benckiser:2011p189}
E.~Benckiser, M.~W. Haverkort, S.~Brück, E.~Goering, S.~Macke,
  A.~Fra{\~{n}}{\'{o}}, X.~Yang, O.~K. Andersen, G.~Cristiani, H.-U.
  Habermeier, A.~V. Boris, I.~Zegkinoglou, P.~Wochner, H.-J. Kim, V.~Hinkov,
  B.~Keimer, Orbital reflectometry of oxide heterostructures.
\newblock {\it Nature Materials\/} {\bf 10}, 189--193 (2011).

\bibitem{Wu:2013p125124}
M.~Wu, E.~Benckiser, M.~W. Haverkort, A.~Frano, Y.~Lu, U.~Nwankwo, S.~Br\"uck,
  P.~Audehm, E.~Goering, S.~Macke, V.~Hinkov, P.~Wochner, G.~Christiani,
  S.~Heinze, G.~Logvenov, H.-U. Habermeier, B.~Keimer, Strain and composition
  dependence of orbital polarization in nickel oxide superlattices.
\newblock {\it Phys. Rev. B\/} {\bf 88}, 125124 (2013).

\bibitem{Poltavets:2010p206403}
V.~V. Poltavets, K.~A. Lokshin, A.~H. Nevidomskyy, M.~Croft, T.~A. Tyson,
  J.~Hadermann, G.~Van~Tendeloo, T.~Egami, G.~Kotliar, N.~ApRoberts-Warren,
  A.~P. Dioguardi, N.~J. Curro, M.~Greenblatt, Bulk magnetic order in a
  two-dimensional {N}i$^{1+}$/{N}i$^{2+}$ (${d}^{9}/{d}^{8}$) nickelate,
  isoelectronic with superconducting cuprates.
\newblock {\it Phys. Rev. Lett.\/} {\bf 104}, 206403 (2010).

\bibitem{Cheng:2012p236403}
J.-G. Cheng, J.-S. Zhou, J.~B. Goodenough, H.~D. Zhou, K.~Matsubayashi,
  Y.~Uwatoko, P.~P. Kong, C.~Q. Jin, W.~G. Yang, G.~Y. Shen, Pressure effect on
  the structural transition and suppression of the high-spin state in the
  triple-layer ${T}^{\ensuremath{'}}$- {L}a$_4${N}i$_3${O}$_8$.
\newblock {\it Phys. Rev. Lett.\/} {\bf 108}, 236403 (2012).

\bibitem{Zhang:2017p864}
J.~Zhang, A.~S. Botana, J.~W. Freeland, D.~Phelan, H.~Zheng, V.~Pardo, M.~R.
  Norman, J.~F. Mitchell, Large orbital polarization in a metallic
  square-planar nickelate.
\newblock {\it Nature Physics\/} {\bf 13}, 864--869 (2017).

\bibitem{Li:2019p624}
D.~Li, K.~Lee, B.~Y. Wang, M.~Osada, S.~Crossley, H.~R. Lee, Y.~Cui, Y.~Hikita,
  H.~Y. Hwang, Superconductivity in an infinite-layer nickelate.
\newblock {\it Nature\/} {\bf 572}, 624--627 (2019).

\bibitem{Hepting:2020p381}
M.~Hepting, D.~Li, C.~Jia, H.~Lu, E.~Paris, Y.~Tseng, X.~Feng, M.~Osada,
  E.~Been, Y.~Hikita, {\it et~al.\/}, Electronic structure of the parent
  compound of superconducting infinite-layer nickelates.
\newblock {\it Nature materials\/} {\bf 19}, 381--385 (2020).

\bibitem{li2020superconducting}
D.~Li, B.~Y. Wang, K.~Lee, S.~P. Harvey, M.~Osada, B.~H. Goodge, L.~F.
  Kourkoutis, H.~Y. Hwang, Superconducting dome in
  ${\mathrm{nd}}_{1\ensuremath{-}x}{\mathrm{sr}}_{x}{\mathrm{nio}}_{2}$
  infinite layer films.
\newblock {\it Phys. Rev. Lett.\/} {\bf 125}, 027001 (2020).

\bibitem{zeng:2020phase}
S.~Zeng, C.~S. Tang, X.~Yin, C.~Li, Z.~Huang, J.~Hu, W.~Liu, G.~J. Omar,
  H.~Jani, Z.~S. Lim, {\it et~al.\/}, Phase diagram and superconducting dome of
  infinite-layer nd$_{1-x}$sr$_x$nio$_2$ thin films.
\newblock {\it arXiv preprint arXiv:2004.11281\/}  (2020).

\bibitem{Botana:2020p011024}
A.~S. Botana, M.~R. Norman, Similarities and differences between
  ${\mathrm{lanio}}_{2}$ and ${\mathrm{cacuo}}_{2}$ and implications for
  superconductivity.
\newblock {\it Phys. Rev. X\/} {\bf 10}, 011024 (2020).

\bibitem{Lechermann:2020p081110}
F.~Lechermann, Late transition metal oxides with infinite-layer structure:
  Nickelates versus cuprates.
\newblock {\it Phys. Rev. B\/} {\bf 101}, 081110 (2020).

\bibitem{Jiang:2019p201106}
P.~Jiang, L.~Si, Z.~Liao, Z.~Zhong, Electronic structure of rare-earth
  infinite-layer $r\mathrm{Ni}{\mathrm{o}}_{2}(r=\mathrm{La},\mathrm{Nd})$.
\newblock {\it Phys. Rev. B\/} {\bf 100}, 201106 (2019).

\bibitem{karp:2020}
J.~Karp, A.~S. Botana, M.~R. Norman, H.~Park, M.~Zingl, A.~Millis, Many-body
  electronic structure of ndnio $ \_2 $ and cacuo $ \_2$.
\newblock {\it arXiv preprint arXiv:2001.06441\/}  (2020).

\bibitem{Hu:2019p032046}
L.-H. Hu, C.~Wu, Two-band model for magnetism and superconductivity in
  nickelates.
\newblock {\it Phys. Rev. Research\/} {\bf 1}, 032046 (2019).

\bibitem{adhikary2020orbital}
P.~Adhikary, S.~Bandyopadhyay, T.~Das, I.~Dasgupta, T.~Saha-Dasgupta, Orbital
  selective superconductivity in a two-band model of infinite-layer nickelates.
\newblock {\it arXiv preprint arXiv:2005.01243\/}  (2020).

\bibitem{Mi:2020p207004}
M.~Jiang, M.~Berciu, G.~A. Sawatzky, Critical nature of the ni spin state in
  doped ${\mathrm{ndnio}}_{2}$.
\newblock {\it Phys. Rev. Lett.\/} {\bf 124}, 207004 (2020).

\bibitem{Zhang2020p023112}
Y.-H. Zhang, A.~Vishwanath, Type-ii $t\text{\ensuremath{-}}j$ model in
  superconducting nickelate
  ${\mathrm{nd}}_{1\ensuremath{-}x}{\mathrm{sr}}_{x}{\mathrm{nio}}_{2}$.
\newblock {\it Phys. Rev. Research\/} {\bf 2}, 023112 (2020).

\bibitem{Zhang:2020p013214}
H.~Zhang, L.~Jin, S.~Wang, B.~Xi, X.~Shi, F.~Ye, J.-W. Mei, Effective
  hamiltonian for nickelate oxides
  ${\mathrm{nd}}_{1\ensuremath{-}x}{\mathrm{sr}}_{x}{\mathrm{nio}}_{2}$.
\newblock {\it Phys. Rev. Research\/} {\bf 2}, 013214 (2020).

\bibitem{Lang:2020doped}
Z.-J. Lang, R.~Jiang, W.~Ku, Where do the doped hole carriers reside in the new
  superconducting nickelates?
\newblock {\it arXiv preprint arXiv:2005.00022\/}  (2020).

\bibitem{karp:2020many}
J.~Karp, A.~S. Botana, M.~R. Norman, H.~Park, M.~Zingl, A.~Millis, Many-body
  electronic structure of ndnio $ \_2 $ and cacuo $ \_2$.
\newblock {\it arXiv preprint arXiv:2001.06441\/}  (2020).

\bibitem{Krishna:2020arxiv}
J.~Krishna, H.~LaBollita, A.~O. Fumega, V.~Pardo, A.~S. Botana, Effects of
  sr-doping on the electronic and spin-state properties of infinite-layer
  nickelates.
\newblock {\it arXiv preprint arXiv:2008.02237\/}  (2020).

\bibitem{Azuma:1992p775}
M.~Azuma, Z.~Hiroi, M.~Takano, Y.~Bando, Y.~Takeda, Superconductivity at 110 k
  in the infinite-layer compound (sr1-xcax)1-ycuo2.
\newblock {\it Nature\/} {\bf 356}, 775--776 (1992).

\bibitem{goodge2020doping}
B.~H. Goodge, D.~Li, M.~Osada, B.~Y. Wang, K.~Lee, G.~A. Sawatzky, H.~Y. Hwang,
  L.~F. Kourkoutis, Doping evolution of the mott-hubbard landscape in
  infinite-layer nickelates.
\newblock {\it arXiv preprint arXiv:2005.02847\/}  (2020).

\bibitem{Motoyama:1996p3212}
N.~Motoyama, H.~Eisaki, S.~Uchida, Magnetic susceptibility of ideal spin 1 $/$2
  heisenberg antiferromagnetic chain systems,
  ${\mathrm{sr}}_{2}{\mathrm{cuo}}_{3}$ and ${\mathrm{srcuo}}_{2}$.
\newblock {\it Phys. Rev. Lett.\/} {\bf 76}, 3212--3215 (1996).

\bibitem{Kojima:1997p1787}
K.~M. Kojima, Y.~Fudamoto, M.~Larkin, G.~M. Luke, J.~Merrin, B.~Nachumi, Y.~J.
  Uemura, N.~Motoyama, H.~Eisaki, S.~Uchida, K.~Yamada, Y.~Endoh, S.~Hosoya,
  B.~J. Sternlieb, G.~Shirane, Reduction of ordered moment and n\'eel
  temperature of quasi-one-dimensional antiferromagnets
  ${\mathrm{sr}}_{2}{\mathrm{cuo}}_{3}$ and
  ${\mathrm{ca}}_{2}{\mathrm{cuo}}_{3}$.
\newblock {\it Phys. Rev. Lett.\/} {\bf 78}, 1787--1790 (1997).

\bibitem{Neudert:1998p657}
R.~Neudert, M.~Knupfer, M.~S. Golden, J.~Fink, W.~Stephan, K.~Penc,
  N.~Motoyama, H.~Eisaki, S.~Uchida, Manifestation of spin-charge separation in
  the dynamic dielectric response of one-dimensional
  ${\mathrm{sr}}_{2}\mathrm{Cu}{O}_{3}$.
\newblock {\it Phys. Rev. Lett.\/} {\bf 81}, 657--660 (1998).

\bibitem{Schlappa:2012p82}
J.~Schlappa, K.~Wohlfeld, K.~Zhou, M.~Mourigal, M.~Haverkort, V.~Strocov,
  L.~Hozoi, C.~Monney, S.~Nishimoto, S.~Singh, {\it et~al.\/}, Spin--orbital
  separation in the quasi-one-dimensional mott insulator sr 2 cuo 3.
\newblock {\it Nature\/} {\bf 485}, 82--85 (2012).

\bibitem{Hirobe:2017p30}
D.~Hirobe, M.~Sato, T.~Kawamata, Y.~Shiomi, K.-i. Uchida, R.~Iguchi, Y.~Koike,
  S.~Maekawa, E.~Saitoh, One-dimensional spinon spin currents.
\newblock {\it Nature Physics\/} {\bf 13}, 30--34 (2017).

\bibitem{Hiroi:1993p315}
.-Z. Hiroi, M.~Takano, M.~Azuma, Y.~Takeda, A new family of copper oxide
  superconductors srn+ 1cuno2n+ 1+ $\delta$ stabilized at high pressure.
\newblock {\it Nature\/} {\bf 364}, 315--317 (1993).

\bibitem{Karmakar:2015p224401}
K.~Karmakar, S.~Singh, Finite-size effects in the quasi-one-dimensional quantum
  magnets
  ${\mathrm{sr}}_{2}{\mathrm{cuo}}_{3},{\mathrm{sr}}_{2}\mathrm{Cu}{}_{0.99}{M}_{0.01}{\mathrm{o}}_{3}(m=\text{Ni},\phantom{\rule{0.16em}{0ex}}\mathrm{Zn}),$
  and ${\mathrm{srcuo}}_{2}$.
\newblock {\it Phys. Rev. B\/} {\bf 91}, 224401 (2015).

\bibitem{Zaanen:1985p418}
J.~Zaanen, G.~A. Sawatzky, J.~W. Allen, Band gaps and electronic structure of
  transition-metal compounds.
\newblock {\it Phys. Rev. Lett.\/} {\bf 55}, 418--421 (1985).

\bibitem{Nimkar:1993p7355}
S.~Nimkar, D.~D. Sarma, H.~R. Krishnamurthy, S.~Ramasesha, Mean-field results
  of the multiple-band extended hubbard model for the square-planar {C}u{O}$_2$
  lattice.
\newblock {\it Phys. Rev. B\/} {\bf 48}, 7355--7363 (1993).

\bibitem{Maiti:1998p1572}
K.~Maiti, D.~D. Sarma, T.~Mizokawa, A.~Fujimori, Electronic structure of
  one-dimensional cuprates.
\newblock {\it Phys. Rev. B\/} {\bf 57}, 1572--1578 (1998).

\bibitem{walters:2009p867}
A.~C. Walters, T.~G. Perring, J.-S. Caux, A.~T. Savici, G.~D. Gu, C.-C. Lee,
  W.~Ku, I.~A. Zaliznyak, Effect of covalent bonding on magnetism and the
  missing neutron intensity in copper oxide compounds.
\newblock {\it Nature Physics\/} {\bf 5}, 867--872 (2009).

\bibitem{karmakar:2015p4843}
K.~Karmakar, R.~Bag, S.~Singh, Crystal growth of spin chain compound sr2cuo3
  doped with quantum defects: Zn, co, ni, and mn.
\newblock {\it Crystal Growth \& Design\/} {\bf 15}, 4843--4853 (2015).

\bibitem{Nucker:1995p8529}
N.~N\"ucker, E.~Pellegrin, P.~Schweiss, J.~Fink, S.~L. Molodtsov, C.~T.
  Simmons, G.~Kaindl, W.~Frentrup, A.~Erb, G.~M\"uller-Vogt, Site-specific and
  doping-dependent electronic structure of
  ${\mathrm{yba}}_{2}$${\mathrm{cu}}_{3}$${\mathrm{o}}_{\mathit{x}}$ probed by
  o 1s and cu 2p x-ray-absorption spectroscopy.
\newblock {\it Phys. Rev. B\/} {\bf 51}, 8529--8542 (1995).

\bibitem{Knupfer:1997pR7291}
M.~Knupfer, R.~Neudert, M.~Kielwein, S.~Haffner, M.~S. Golden, J.~Fink, C.~Kim,
  Z.-X. Shen, M.~Merz, N.~N\"ucker, S.~Schuppler, N.~Motoyama, H.~Eisaki,
  S.~Uchida, Z.~Hu, M.~Domke, G.~Kaindl, Site-specific unoccupied electronic
  structure of one-dimensional ${\mathrm{srcuo}}_{2}$.
\newblock {\it Phys. Rev. B\/} {\bf 55}, R7291--R7294 (1997).

\bibitem{Neudert:2000p10752}
R.~Neudert, S.-L. Drechsler, J.~M\'alek, H.~Rosner, M.~Kielwein, Z.~Hu,
  M.~Knupfer, M.~S. Golden, J.~Fink, N.~N\"ucker, M.~Merz, S.~Schuppler,
  N.~Motoyama, H.~Eisaki, S.~Uchida, M.~Domke, G.~Kaindl, Four-band extended
  hubbard hamiltonian for the one-dimensional cuprate
  ${\mathrm{sr}}_{2}{\mathrm{cuo}}_{3}:$ distribution of oxygen holes and its
  relation to strong intersite coulomb interaction.
\newblock {\it Phys. Rev. B\/} {\bf 62}, 10752--10765 (2000).

\bibitem{Chakhalian:2007p1114}
J.~Chakhalian, J.~W. Freeland, H.-U. Habermeier, G.~Cristiani, G.~Khaliullin,
  M.~van Veenendaal, B.~Keimer, Orbital reconstruction and covalent bonding at
  an oxide interface.
\newblock {\it Science\/} {\bf 318}, 1114-1117 (2007).

\bibitem{Moretti_Sala_2011}
M.~M. Sala, V.~Bisogni, C.~Aruta, G.~Balestrino, H.~Berger, N.~B. Brookes,
  G.~M. de~Luca, D.~D. Castro, M.~Grioni, M.~Guarise, P.~G. Medaglia, F.~M.
  Granozio, M.~Minola, P.~Perna, M.~Radovic, M.~Salluzzo, T.~Schmitt, K.~J.
  Zhou, L.~Braicovich, G.~Ghiringhelli, Energy and symmetry of dd excitations
  in undoped layered cuprates measured by {CuL}3resonant inelastic x-ray
  scattering.
\newblock {\it New Journal of Physics\/} {\bf 13}, 043026 (2011).

\bibitem{Ushakov:2011p445601}
A.~V. Ushakov, S.~V. Streltsov, D.~I. Khomskii, Crystal field splitting in
  correlated systems with negative charge-transfer gap.
\newblock {\it Journal of Physics: Condensed Matter\/} {\bf 23}, 445601 (2011).

\bibitem{Disa:2015p026801}
A.~S. Disa, D.~P. Kumah, A.~Malashevich, H.~Chen, D.~A. Arena, E.~D. Specht,
  S.~Ismail-Beigi, F.~J. Walker, C.~H. Ahn, Orbital engineering in
  symmetry-breaking polar heterostructures.
\newblock {\it Phys. Rev. Lett.\/} {\bf 114}, 026801 (2015).

\bibitem{Liao_2019}
Z.~Liao, E.~Skoropata, J.~W. Freeland, E.-J. Guo, R.~Desautels, X.~Gao,
  C.~Sohn, A.~Rastogi, T.~Z. Ward, T.~Zou, T.~Charlton, M.~R. Fitzsimmons,
  H.~N. Lee, Large orbital polarization in nickelate-cuprate heterostructures
  by dimensional control of oxygen coordination.
\newblock {\it Nature Communications\/} {\bf 10} (2019).

\bibitem{Peil:2014p045128}
O.~E. Peil, M.~Ferrero, A.~Georges, Orbital polarization in strained
  {L}a{N}i{O}$_3$: Structural distortions and correlation effects.
\newblock {\it Phys. Rev. B\/} {\bf 90}, 045128 (2014).

\bibitem{Middey:2016p056801}
S.~Middey, D.~Meyers, D.~Doennig, M.~Kareev, X.~Liu, Y.~Cao, Z.~Yang, J.~Shi,
  L.~Gu, P.~J. Ryan, R.~Pentcheva, J.~W. Freeland, J.~Chakhalian, Mott
  electrons in an artificial graphenelike crystal of rare-earth nickelate.
\newblock {\it Phys. Rev. Lett.\/} {\bf 116}, 056801 (2016).

\bibitem{Stavitski:2010p687}
E.~Stavitski, F.~M. De~Groot, The ctm4xas program for eels and xas spectral
  shape analysis of transition metal l edges.
\newblock {\it Micron\/} {\bf 41}, 687--694 (2010).

\bibitem{retegan_crispy}
M.~Retegan, Crispy: v0.7.3 (2019).

\bibitem{Utz_2017}
Y.~Utz, F.~Hammerath, R.~Kraus, T.~Ritschel, J.~Geck, L.~Hozoi, J.~van~den
  Brink, A.~Mohan, C.~Hess, K.~Karmakar, S.~Singh, D.~Bounoua, R.~Saint-Martin,
  L.~Pinsard-Gaudart, A.~Revcolevschi, B.~Büchner, H.-J. Grafe, Effect of
  different in-chain impurities on the magnetic properties of the spin chain
  compound {SrCuO}2 probed by {NMR}.
\newblock {\it Physical Review B\/} {\bf 96} (2017).

\bibitem{van_der_Laan_1988}
G.~van~der Laan, B.~T. Thole, G.~A. Sawatzky, M.~Verdaguer, Multiplet structure
  in the ${L}_{2}$,3 x-ray-absorption spectra: A fingerprint for high- and
  low-spin ${\mathrm{ni}}^{2+}$ compounds.
\newblock {\it Phys. Rev. B\/} {\bf 37}, 6587--6589 (1988).

\bibitem{calc}
$\Delta$ = 9 eV, $U_{dd}$ = 7 eV, $U_{pd}$ = 9 eV were used in Cluster
  calculations. The hopping parameters, related to $d_{z^2}$, $d_{x^2 - y^2}$,
  $d_{xy}$ and $d_{xz}$/$d_{yz}$ orbitals, were chosen to be 2.0, 2.0, 1.0, 1.0
  eV, respectively. For the high spin ($S$ = 1) configuration, the crystal
  field parameters for $D_{4h}$ symmetry were chosen as $Dq$ = 0.235 eV, $Ds$ =
  0.1 eV, $Dt$ = 0.05 eV. Whereas, $Dq$ = 0.235 eV, $Ds$ = 0.5eV, $Dt$ = 0.3 eV
  were used for the low spin ($S$ = 0) configuration.

\bibitem{Johnston:2000p9558}
D.~C. Johnston, R.~K. Kremer, M.~Troyer, X.~Wang, A.~Kl\"umper, S.~L. Bud'ko,
  A.~F. Panchula, P.~C. Canfield, Thermodynamics of spin $s=1/2$
  antiferromagnetic uniform and alternating-exchange heisenberg chains.
\newblock {\it Phys. Rev. B\/} {\bf 61}, 9558--9606 (2000).

\bibitem{Fujimoto:2004p037206}
S.~Fujimoto, S.~Eggert, Boundary susceptibility in the spin-$1/2$ chain:
  Curie-like behavior without magnetic impurities.
\newblock {\it Phys. Rev. Lett.\/} {\bf 92}, 037206 (2004).

\bibitem{Sirker:2007p137205}
J.~Sirker, N.~Laflorencie, S.~Fujimoto, S.~Eggert, I.~Affleck, Chain breaks and
  the susceptibility of
  ${\mathrm{sr}}_{2}{\mathrm{cu}}_{1\ensuremath{-}x}{\mathrm{pd}}_{x}{\mathrm{o}}_{3+\ensuremath{\delta}}$
  and other doped quasi-one-dimensional antiferromagnets.
\newblock {\it Phys. Rev. Lett.\/} {\bf 98}, 137205 (2007).

\bibitem{Ami:1995p5994}
T.~Ami, M.~K. Crawford, R.~L. Harlow, Z.~R. Wang, D.~C. Johnston, Q.~Huang,
  R.~W. Erwin, Magnetic susceptibility and low-temperature structure of the
  linear chain cuprate ${\mathrm{sr}}_{2}$${\mathrm{cuo}}_{3}$.
\newblock {\it Phys. Rev. B\/} {\bf 51}, 5994--6001 (1995).

\bibitem{van1932book}
J.~H. Van~Vleck, {\it The theory of electric and magnetic susceptibilities\/}
  (Clarendon Press, 1932).

\bibitem{Maiti:1999p12457}
K.~Maiti, P.~Mahadevan, D.~D. Sarma, Evolution of electronic structure with
  dimensionality in divalent nickelates.
\newblock {\it Phys. Rev. B\/} {\bf 59}, 12457--12470 (1999).

\bibitem{Giannozzi:2009p395502}
P.~Giannozzi, S.~Baroni, N.~Bonini, M.~Calandra, R.~Car, C.~Cavazzoni,
  D.~Ceresoli, G.~L. Chiarotti, M.~Cococcioni, I.~Dabo, A.~D. Corso,
  S.~de~Gironcoli, S.~Fabris, G.~Fratesi, R.~Gebauer, U.~Gerstmann,
  C.~Gougoussis, A.~Kokalj, M.~Lazzeri, L.~Martin-Samos, N.~Marzari, F.~Mauri,
  R.~Mazzarello, S.~Paolini, A.~Pasquarello, L.~Paulatto, C.~Sbraccia,
  S.~Scandolo, G.~Sclauzero, A.~P. Seitsonen, A.~Smogunov, P.~Umari, R.~M.
  Wentzcovitch, {QUANTUM} {ESPRESSO}: a modular and open-source software
  project for quantum simulations of materials.
\newblock {\it Journal of Physics: Condensed Matter\/} {\bf 21}, 395502 (2009).

\bibitem{Perdew:1996p3865}
J.~P. Perdew, K.~Burke, M.~Ernzerhof, Generalized gradient approximation made
  simple.
\newblock {\it Phys. Rev. Lett.\/} {\bf 77}, 3865--3868 (1996).

\bibitem{Vanderbilt:1990p7892}
D.~Vanderbilt, Soft self-consistent pseudopotentials in a generalized
  eigenvalue formalism.
\newblock {\it Physical Review B\/} {\bf 41}, 7892--7895 (1990).

\bibitem{Cococcioni:2005p3}
M.~Cococcioni, S.~de~Gironcoli, Linear response approach to the calculation of
  the effective interaction parameters in the $\mathrm{LDA}+\mathrm{U}$ method.
\newblock {\it Phys. Rev. B\/} {\bf 71}, 035105 (2005).

\end{thebibliography}

\section*{Acknowledgement}
 SM thanks  Tanmoy Das and H. R. Krishnamurthy for discussions.  SM acknowledges financial support from a SERB Early Career Research Award (ECR/2018/001512) and MHRD STARS 156/2019 project for research funding.  SM also thanks the Department of Science and Technology, India (No. SR/NM/Z-07/2015) for the financial support to conduct the synchrotron experiment at the Advanced Photon Source and Jawaharlal Nehru Centre for Advanced Scientific Research (JNCASR) for managing the project. This research used resources of the Advanced Photon Source, a U.S. Department of Energy Office of Science User Facility operated by Argonne National Laboratory under Contract No. DE-AC02-06CH11357. AN acknowledges support from the start-up grant (SG/MHRD-19-0001) of the Indian Institute of Science.



\noindent

\end{document}